\documentstyle[aps,epsfig]{revtex}
\input epsf

\def\rmmat#1{{\hbox{\rm #1}}}
\def\rmscr#1{\rmmat{\scriptsize #1}}
\newcommand{\be}{\begin{equation}}
\newcommand{\ee}{\end{equation}}
\newcommand{\ba}{\begin{eqnarray}}
\newcommand{\ea}{\end{eqnarray}}
\def\eqref#1{Eq.~\ref{eq:#1}}

\def\hOm{{\hat\Omega}}
\def\rmmat#1{{\hbox{\rm #1}}}
\def\rmscr#1{\rmmat{\scriptsize #1}}

\begin{document}
\title{QED and the High Polarization of the 
       Thermal Radiation from Neutron Stars}
\author{Jeremy S. Heyl\footnote{Chandra Fellow}}
\address{
       Harvard-Smithsonian Center for Astrophysics, MS-51,
       60 Garden Street, Cambridge, Massachusetts 02138, United States}
\author{Nir J. Shaviv\footnote{Current Address: Racah Institute of Physics, Hebrew University
       Giv'at Ram, Jerusalem 91904, Israel}}
\address{
       Canadian Institute for Theoretical
       Astrophysics, University of Toronto, 60 St. George St., 
       Toronto, ON M5S 3H8, Canada
}
\maketitle
\begin{abstract}
The thermal emission of strongly magnetized neutron-star atmospheres
is thought to be highly
polarized\cite{Kann75,polaratmos1,polaratmos2}.  However, because of
the different orientations of the magnetic field over the surface of
the neutron star (NS), it is commonly assumed that the net observed
polarization will be significantly reduced as the polarization from
different regions will cancel each other\cite{Kann75,Pavl00Thermal}.
We show that the birefringence of the magnetized QED
vacuum\cite{QEDbire1,QEDbire2} decouples the polarization modes in
the magnetosphere; therefore, the direction of the polarization
follows the direction of the magnetic field up to a large
distance\cite{Heyl00Polar} from the stellar surface.  At this
distance, the rays that leave the surface and are destined for our
detectors pass through only a small solid angle; consequently, the
polarization direction of the emission originating in different
regions will tend to align together.  The net observed polarization
of the thermal radiation of NSs should therefore be very large.
Measurement of this polarization will be the first direct evidence
of the birefringence of the magnetized vacuum due to QED and a
direct probe of behavior of the vacuum at magnetic fields of order
of and above the critical QED field of $4.4\times 10^{13}$~G.  The
large observable polarization will also help us learn more about the
atmospheric properties of NSs.
\end{abstract}

\section{Introduction}

 When strong magnetic fields are present, the atmospheres of NSs have
 a significantly different opacity in the two polarization
 eigenmodes\cite{1974ApJ...190..141L,1979PhRvD..19.1684V}.  As a
 result, the effective depth to which an observer will be able to see
 the atmosphere in the two polarizations will be significantly
 different, and each one will therefore have a different effective
 temperature.  The net effect is that each element on the NS surface
 emits highly polarized ($>50\%$) thermal radiation at optical through
 X-ray frequencies\cite{Kann75,polaratmos1,polaratmos2}; at the point
 of emission, the direction of polarization is correlated with the
 direction of the magnetic field.  However, the magnetic field
 orientation varies over the surface of a neutron star.  When the
 polarized intensities are then summed, a relatively small net
 polarization results.  Typical values of 5\% to 25\% are
 obtained\cite{Kann75,Pavl00Thermal}.  This however neglects the
 effects that QED has on the propagation of photons through the
 vacuum.

 Heyl \& Shaviv \cite{Heyl00Polar} examined the specific case of
 purely radial photon trajectories from a rotating neutron star.
 Since each line of sight is represented by a single trajectory,
 no depolarization occurs.  The net polarization observed is
 guaranteed to be equal in magnitude to that produced.  In this paper,
 we examine the complementary issue of how photons propagate near the
 surface of the star.  In this case both non-radial trajectories and
 general relativity have important effects as seen in earlier
 work\cite{Pavl00Thermal}.  However, no previous work has included
 a realistic treatment of the optically active magnetosphere of a
 neutron star\cite{Heyl00Polar} to examine the extent of
 depolarization of the thermal radiation from the surface of a neutron
 star.

\section{Physical Background}

 In the presence of strong magnetic fields, QED renders the vacuum
 birefringent -- the indices of refraction of the two linear
 polarization modes differ from each other\cite{QEDbire1,QEDbire2}.
 In spite of over a century of effort, the effect of a magnetic field
 on the speed of light in vacua has not been detected
 \cite{Iaco79,Baka98,Rizz98,Nezr99}.

 A convenient formalism to describe the birefringence of a medium and
 the effects it has on the polarization states is by using the
 Poincar\'e space. The latter can describe the polarization state
 vector ${\bf s}$ (which is the normalized Stokes vector) and a
 complex linear combinations over the vector space perpendicular to
 the light ray \cite{Poincare_Comment}.

 In Poincar\'e space, the birefringence can be described by a vector
 $\hOm$ which points in the direction of the faster moving
 polarization mode \cite{Poincare_Comment} and of which the amplitude
 is the difference between the wavenumbers of the two modes for a
 particular frequency ($\Delta k$).  The faster mode is polarized
 perpendicular to the direction of ${\bf B}_\perp$.  For weak fields
 $B \ll B_{QED}=4.4\times 10^{13}$~G, Heyl and
 Shaviv\cite{Heyl00Polar} find
\begin{equation}
|\hOm| = \frac{\alpha}{15} \frac{\nu}{c} 
\left ( \frac{B_\perp}{B_{QED}} \right)^2
\end{equation}
 where $B_\perp$ is the strength of the magnetic field perpendicular
 to the propagation direction of the photon, $\nu$ is the photon
 frequency and $c$ is the speed of light in vacuum.  The origin of the
 vector $\hOm$ is purely a QED effect arising from the interaction of
 the outgoing photons with the virtual electrons of the vacuum. A
 similar birefringent vector arises when the interaction with the
 plasma electrons is taken into account, however, this will be
 important only at frequencies below the optical\cite{Chen79,Barn86}.
 
 Using the result of Kubo and Nagata\cite{Kubo83} to describe the
 polarization evolution in a dielectric birefringent medium, Heyl and
 Shaviv\cite{Heyl00Polar} have shown that this vacuum birefringence
 will couple the evolution of the photon polarization to the magnetic
 field direction through the equation:
\begin{equation}
 {\partial {\bf s} \over \partial x_3} = \hOm \times {\bf s},
\label{eq:s}
\end{equation}
 where $x_3$ is the distance along the direction of propagation. ${\bf
 s}$ is the normalized Stokes vector, and $\hOm$ is the birefringent
 vector previously described.  

 Close to the neutron star, $\hOm$ is large, and ${\bf s}$ rotates
 quickly around $\hOm$.  Thus, as the direction of the magnetic field
 changes, $\hOm$ will rotate and ${\bf s}$ will follow it
 adiabatically. Far enough from the NS, the amplitude of $\hOm$ will
 fall and ${\bf s}$ will not be able to follow $\hOm$ any longer. The
 condition for adiabaticity to hold is \cite{Heyl00Polar}.  
\begin{equation}
l ( \Delta k ) = \left| \hOm / (\nabla \ln
 \left| \hOm \right| ) \right| \gtrsim 0.5 
\end{equation}
 where $l$ is the scale length of the magnetic field.

 If one assumes that the field surrounding the star has a dipolar
 geometry, the polarization states evolve adiabatically if 
\begin{eqnarray}
 r \lesssim r_\rmscr{pl} &\equiv& \left ( \frac{\alpha}{45}
 \frac{\nu}{c} \right )^{1/5} \left ( \frac{\mu}{B_\rmscr{QED}} \sin
 \beta \right )^{2/5} \\ \nonumber &\approx & 1.2 \times 10^{7} \left
 ( \frac{\mu}{10^{30}~\rmmat{G cm}^3} \right )^{2/5} \left (
 \frac{\nu}{10^{17}~\rmmat{Hz}} \right)^{1/5} \left ( \sin \beta
 \right)^{2/5} \rmmat{cm},
\end{eqnarray}
 where $r$ is the distance from the center of the star, $\mu$ is the
 magnetic dipole moment of the neutron star, and $\beta$ is the angle
 between the dipole axis and the line of sight; $r_\rmscr{pl}$ is the
 polarization-limiting radius, borrowing terminology from the study of
 radio pulsars \cite{Chen79}.

 Physically, the adiabatic regime is appropriate as long as $\Delta n
 \cdot k \cdot l \gtrsim 1$, where $\Delta n$ is the difference
 between the indices of refraction, $k$ is the wavevector and $l \sim
 r$ is the the distance scale over which the physical variables
 change. In other words, adiabaticity requires that over the physical
 length scale of the problem, the two modes develop a significant
 phase difference between them. The coupling or polarization limiting
 radius therefore does not depend on the rate at which $\hOm$ changes
 direction. The direction (and rate of change in direction) of $\hOm$
 will however determine the polarization left beyond $r_\rmscr{pl}$. Since
 $\hOm$ is in the $1-2$ plane describing linear polarizations, the
 direction of $\hOm$ at $r_\rmscr{pl}$ will determine the linear
 polarization component of ${\bf s}$ while the rate of change of
 ${\hOm}$ will determine the circular component.

 Heuristically, because of vacuum birefringence in the magnetosphere,
 the observed polarization direction from a surface element is
 correlated with the direction of the magnetic field far from the
 stellar surface.  At this distance, the bundle of rays that will
 eventually be detected passes through a small solid angle.  Over this
 small solid angle, the direction of the magnetic field varies little,
 so the observed polarization from different parts also varies little
 and a large net polarization results.  Furthermore, this heuristic
 picture predicts that because stars with smaller radii generally
 result in smaller ray-bundles, smaller stars will exhibit a larger
 net polarization.

 This heuristic picture is bourn out by detailed calculations.  To
 calculate the process accurately, one calculates the photon
 trajectory both in spacetime and on the Poincar\'e sphere in the
 context of general relativity (GR).  First, we incorporate light
 bending (as is given for example by Page\cite{Page95}).  Second, we
 have to use the result of Pineault\cite{Pine77} who showed that along
 the bent light rays present in GR, the polarization direction rotates
 in such a way that it keeps fixed orientation with respect to the
 normal to the trajectory plane, remaining perpendicular to the
 wavevector.  Additionally, GR distorts the dipole magnetic
 field\cite{Ginz65} near the star.

\section{Results}

 A sample result of the integration of the polarization is given in
 Fig.~1, where the apparent polarization at infinity is depicted
 together with the direction that the polarization would have had if
 the effect of QED polarization alignment would have been artificially
 switched off.  Clearly, the effect of the adiabatic evolution of
 polarization close to the neutron star is to align the polarization
 vectors such that they sum mostly coherently. A large net
 polarization could therefore be maintained.  Moreover, when the modes
 couple close to the the neutron star (for low frequencies or weak
 magnetic fields) a circular component of the polarization is
 generated.  However, this component averages to zero if the magnetic
 field has cylindrical symmetry (for example, if only a centered
 magnetic dipole moment exists).

 If we wish to find the net polarization that an observer will
 measure, we need to know the intensity as a function of angle and
 energy of each surface element.  For that, detailed atmospheric
 models are needed; this is beyond the scope of this article.
 Instead, we will simplify the process by assuming that all surface
 elements emit radiation isotropically (into their upper hemisphere),
 with the same flux, and only in one polarization.  In real
 atmospheres, the emission is angle dependent and surface elements
 closer to the poles are somewhat hotter and thus emit more radiation.
 The emission from a surface element is not completely polarized but
 polarized fractions greater than 50\% are typical
 \cite{polaratmos1,polaratmos2}.  We neglect these complications and
 defer them to a more detailed analysis since we mainly wish to show
 at this point that QED effects are important.  The inclusion of a
 detailed atmosphere still results with the same conclusion that a
 high net polarization is obtained\cite{Heyl01polar}.

 The results when assuming an isothermal surface with an isotropic
 emission are given in Fig.~2.  Plotted are the net observed
 polarized fractions as a function of frequency for two different
 angles between the line of sight and the dipole axis,
 $\beta=30^\circ$ and $\beta=60^\circ$, each assuming
 three different radii for the neutron star.  $\mu_{30}$ is the
 magnetic dipole moment of the NS measured in units of $10^{30}~{\rm
 G~ cm^{3}}$ to which the frequency is normalized.

 The extent of polarization increases dramatically with increasing
 frequency which results from the increase in $r_\rmscr{pl}$ with
 energy.  At low energies, QED is unimportant and we recover the
 earlier result \cite{Pavl00Thermal} that more compact stars (smaller
 values of $R/M$) exhibit less polarization.  A distant observer sees
 a larger fraction of the surface of a more compact star due to the
 general relativistic bending of the photon trajectories.

 At higher energies where
 $r_\rmscr{pl} \gg\ r$, the opposite trend is evident.  Since stars
 with larger radii subtend at larger solid angle at $r_\rmscr{pl}$,
 the extent of polarization decreases as the stellar radius increases.
 This is the opposite of the trend expected from GR alone.
 Additionally, the extent of the polarization also depends on the
 angle between the line of sight and the dipole axis ($\beta$), which
 reflects the dependence of $r_\rmscr{pl}$ on $\beta$.

\section{Discussion}

 QED has various effects on the emission from neutron stars. For example, it was
 shown that is should be properly taken into account when calculating
 the atmospheric opacity \cite{polaratmos1,Mesz79}. Here, it was shown
 that QED has a startling effect on the polarization of photons while
 traversing the neutron-star magnetosphere.  Gnedin, Pavlov and Shibanov
 \cite{1978SvAL....4..117G} noted that QED may be important in
 reprocessing radiation in accreting neutron stars.

 The intrinsic polarization of the radiation emitted thermal from the
 surface of a neutron star with a magnetic field stronger than a few
 times $10^{11}$~G is very high
 \cite{Kann75,polaratmos1,polaratmos2,Heyl01polar} (50\% to 80\% in
 optical through X-ray bands). Thus, at the peak of their thermal
 emission $\sim 10^{18}$ Hz, average pulsars should be highly
 polarized.  Even at optical wavelengths, their net polarization
 should be significantly higher due to QED alignment.  In the optical,
 the extent of the polarization is strongly sensitive to the angle
 that the magnetic dipole makes with the line of sight.  Furthermore,
 in agreement with the heuristic picture presented earlier, stars with
 smaller radii (but the same magnetic dipole moment) exhibit larger
 net polarizations.  The polarized fraction in the optical increases
 by nearly a factor of two as the radius of the star shrinks from
 18~km down to 6~km.
 
 Currently, optical or UV polarimetry of neutron stars could be
 obtained in principle. However, since the thermal radiation of even
 the closest neutron star is very faint, it requires very long exposures on even
 the largest instruments available.  In the long run, X-ray
 polarimetry will probably be more favorable since the thermal
 emission peaks at X-rays.

 Since the intrinsic polarization is not wiped out, more information
 can be extracted about neutron stars and their atmospheres.  This may allow the
 measurement of $B$, $M/R$ and possibly the basic composition of their
 atmosphere which affects the spectral and polarization properties
 (for example, whether they are rich in H, He or Fe).

 The last experiment to have X-ray polarimetry capability was more
 than 20 years ago aboard the OSO-8 satellite\cite{Silv78,Hugh84}.  
 Clearly, a new X-ray polarimeter will be most rewarding and is long
 overdue\cite{Soff97,Soff00,Cost01}.  The measurement of strong polarization
 could prove fundamental QED physics in the extreme-field regime.

\begin{acknowledgements}
Support for JSH was provided by the National Aeronautics and
Space Administration through Chandra Postdoctoral Fellowship Award
Number PF0-10015 issued by the Chandra X-ray Observatory Center, which
is operated by the Smithsonian Astrophysical Observatory for and on
behalf of NASA under contract NAS8-39073.
NS Wishes to thank CITA for
the fellowship supporting him and the Racah Institute of Physics,
Hebrew University, for Hosting him while some of this work was done.
\end{acknowledgements}

\begin{figure}
{\epsfig{file=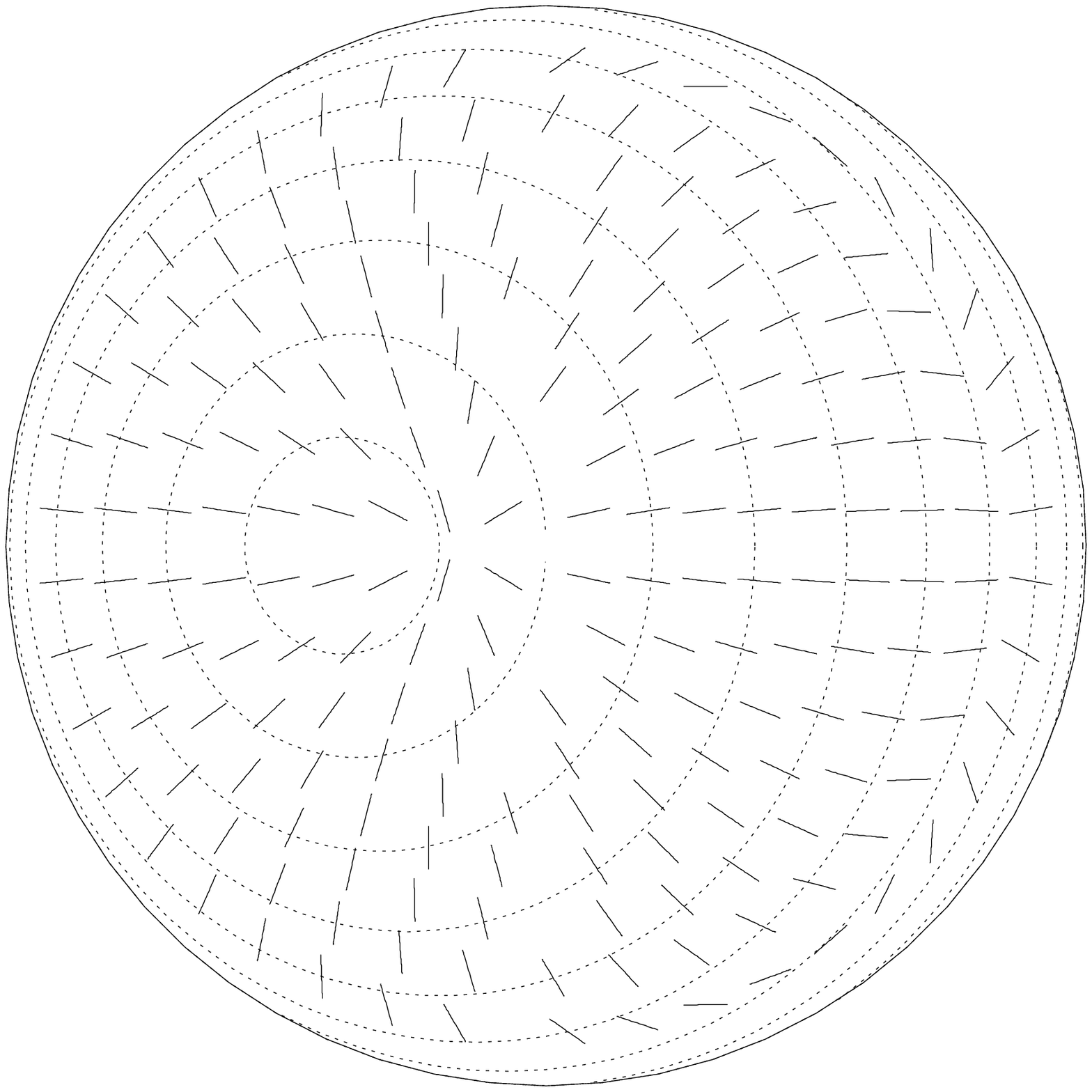,width=3in}
\epsfig{file=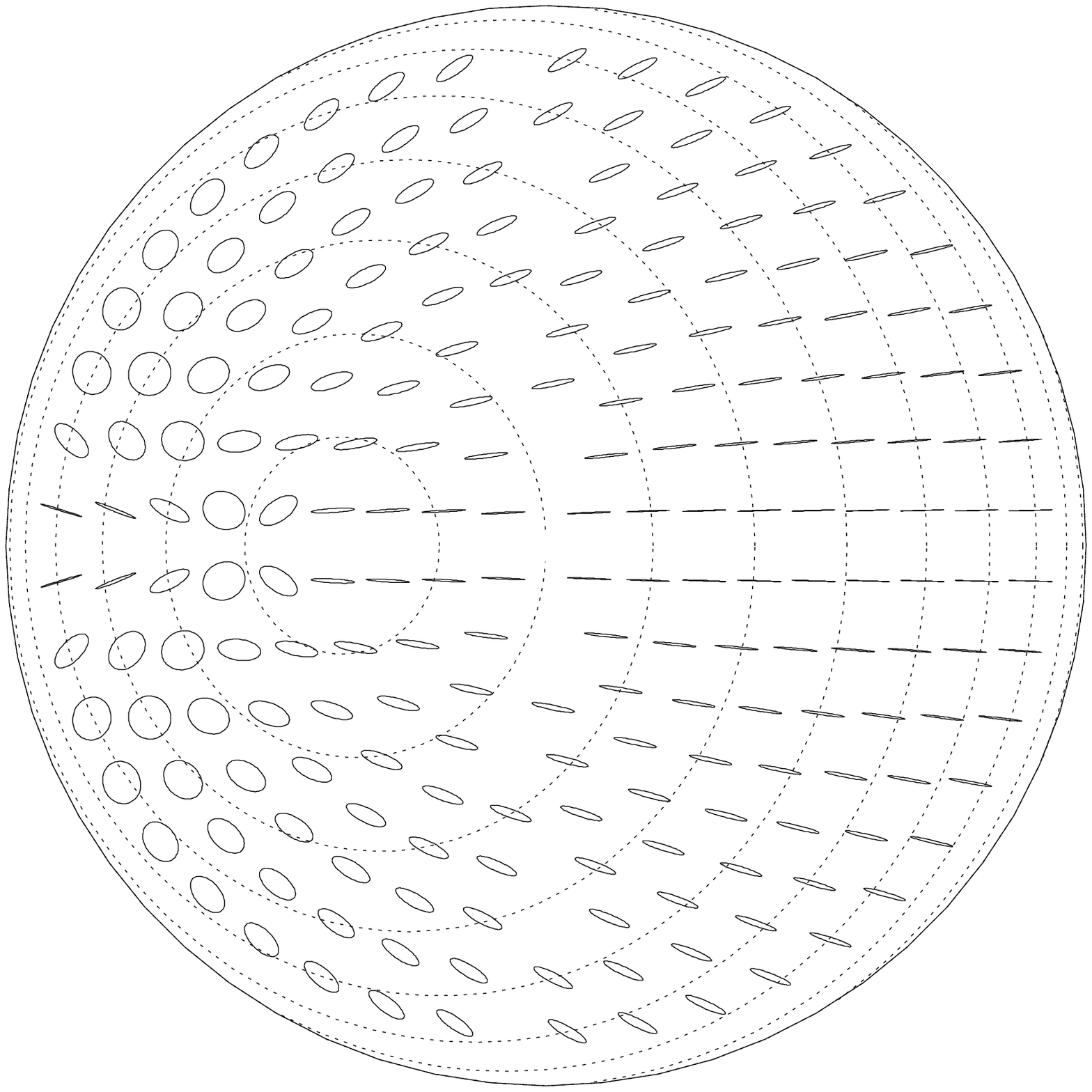,width=3in}}
  \caption{ The apparent polarized image of a neutron star overlayed
  on the apparent image of the NS.  The left panel depicts the
  observed map of polarization directions if one assumes that the
  surface emits only in the ordinary mode and neglects the vacuum
  birefringence induced by QED.  The right panel shows the
  polarization map including birefringence for a frequency of $\nu =
  \mu_{30}^{-2} 10^{17} Hz$. The ellipses and short lines describe the
  polarization of a light ray originating from the surface element
  beneath them. The lines and the major axes of the ellipses point
  towards the direction of the linear component of the polarization
  direction. The minor to major axis ratio provides the amount of
  circular polarization ($s_3$).  In both maps, the large dashed
  curves are lines of constant magnetic latitude (separated by
  $15^{\circ}$). The observer's line of sight makes an angle of
  $30^\circ$ with the dipole axis.  For comparison, the net linear
  polarization on the left 13\% while it is 70\% on the right. In a
  more realistic NS, the values for X-ray frequencies would be reduced
  to 6-10\% and 35-55\% respectively [21]
  since the intrinsic polarization of
  each element is not 100\% but smaller.
}
\end{figure}
\clearpage
\begin{figure}
\centerline{\epsfig{file=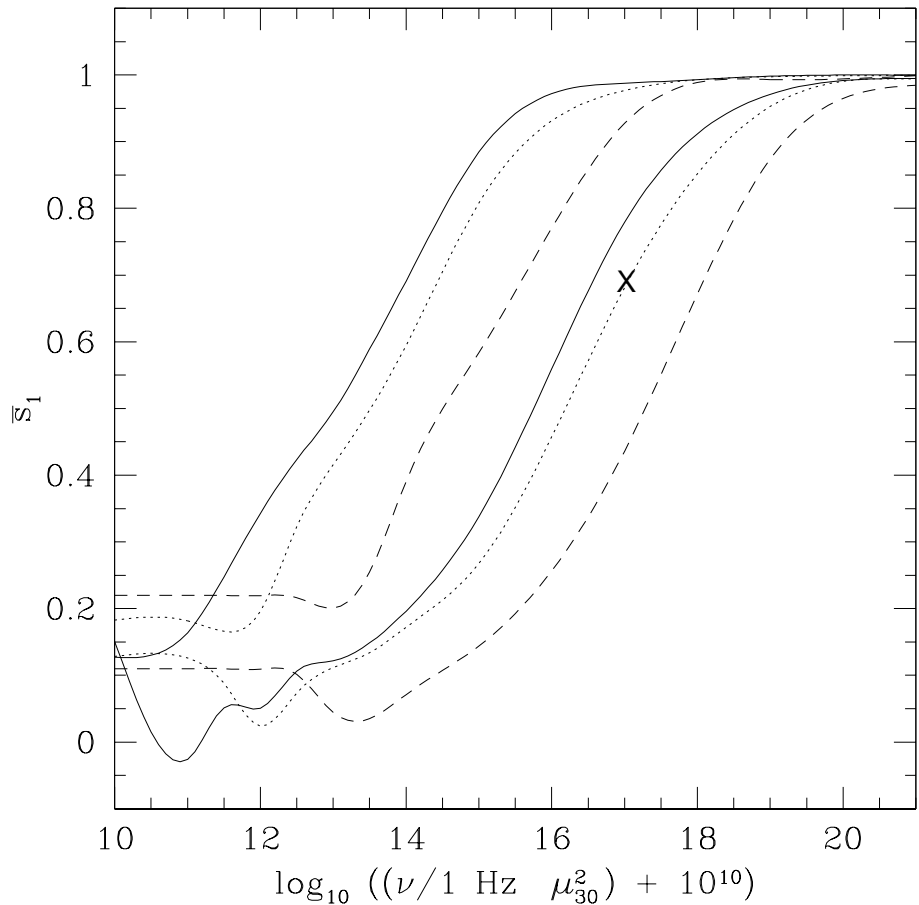,width=5in}} 
  \caption{ The net polarization to be observed as a function of
  frequency for three different NS radii (solid line -- 6 km, dotted
  -- 10 km, and dashed -- 18 km) and two observer magnetic
  co-latitudes (upper three curves -- $\beta=60^{\circ}$, lower three
  curves -- $\beta=30^{\circ}$).  The graphs assume that 
  the surface has a uniform temperature and the emissivity is
  spherically symmetric. The case depicted in the previous figure is
  marked by an ``X''. It should be compared with the low frequency
  limit of the curve, for which QED is unimportant. }
\end{figure}

\bibliographystyle{prsty}

 \def\PoincareComment{The 1-2 plane of the Poincar\'e sphere of
 polarizations describes linearly polarized states with a rotational
 symmetry that is half of real space (namely, a full rotation in the
 1-2 plane corresponds to half a rotation in real space, in the plane
 perpendicular to the light ray). The poles (the $\pm 3$-directions)
 describe circular polarizations. Intermediate directions are
 elliptically polarized. Values of $|{\bf s}|<1$ describe less
 coherent polarization states. The +1 direction can be chosen
 arbitrarily to describe a linear polarizations in the direction of
 the projection of the magnetic axis onto the plane perpendicular to
 the light ray, such that the -1 direction describes the perpendicular
 polarization state and the $\pm 2$ directions describe polarizations
 which are $\pm 45^{\circ}$ away from the apparent magnetic axis. The
 `direction of the magnetic field' (or any other vector) in Poincar\'e
 space, is the direction in Poincar\'e space that corresponds to light
 polarized in the direction of the projection of ${\bf B}$ (or any
 other vector) in real space, onto the plane perpendicular to the
 light ray.}

\bibliography{polarpaps}

\end{document}